\documentclass[fleqn,twoside]{article}
\usepackage[headings]{espcrc2}

\readRCS $Id: espcrc2.tex,v 1.2 2004/02/24 11:22:11 spepping Exp $ \usepackage{graphicx,epsfig}
\newcommand{\selen}{$^{74}$Se}
\newcommand{\german}{$^{74}$Ge}
\newcommand{\nnbbm}{$2\nu\beta^-\beta^-$}
\newcommand{\bbm}{$\beta^-\beta^-$}
\newcommand{\nnbbp}{$2\nu\beta^+\beta^+$}
\newcommand{\nnbec}{$2\nu\beta^+\varepsilon$}
\newcommand{\nnecec}{$2\nu\varepsilon\varepsilon$}
\newcommand{\dbb}{$\beta\beta$}
\newcommand{\znbbm}{$0\nu\beta^-\beta^-$}
\newcommand{\znbbp}{$0\nu\beta^+\beta^+$}
\newcommand{\znbec}{$0\nu\beta^+\varepsilon$}
\newcommand{\znecec}{$0\nu\varepsilon\varepsilon$}
\newcommand{\ecec}{$\varepsilon\varepsilon$}

\begin{document}

\title{Nuclear-atomic state degeneracy
in neutrinoless double-electron capture:\\ \centerline{A unique test for a Majorana-neutrino}}

\author{
D. Frekers
\address[IKP]{Institut f\"ur Kernphysik, Westf\"alische Wilhelms-Universit\"at M\"unster, D-48149 M\"unster, Germany}
}

\runtitle{Nuclear-atomic state degeneracy in neutrinoless double-electron capture} \runauthor{D.Frekers}

\begin{abstract}
There is a general consensus that detection of a double-beta ($\beta\beta$) decay without a neutrino involved
would mark physics beyond the Standard Model. This is because lepton number conservation would be violated and
the neutrino would reveal itself as being its own antiparticle, thereby of Majorana type. So
far, the experimental focus has mostly been on the \bbm\ decay variant, where one attempts to measure the
spectrum of the two emitted electrons. A discrete line at the endpoint energy marks the signature for a Majorana
neutrino. Little attention has been given to alternative decay modes in \dbb\ decay. In this note we show
that there is at least one case in the periodic table, where the parent in the neutrinoless double-electron
capture process (\znecec\ decay) is nearly degenerate with an excited state in the daughter, leading to a
possible enhancement of the decay rate by several orders of magnitude. It is the nucleus \selen, which has this
remarkable property. Furthermore, there is an easy to detect $2\gamma$ ray decay cascade in \german, which
follows the \znecec, and which by its mere detection provides unique signature of the Majorana neutrino.

\end{abstract}

\maketitle

\section{Introduction}

Measuring any properties of the neutrino is a formidable experimental challenge. It took e.g. more than 40 years
after its discovery to prove that the neutrino is massive~\cite{Super-K}, though it has still not been possible
to assign an absolute value to the mass. To prove, whether the neutrino is a Dirac or a Majorana particle is another
challenge that still lies ahead, and rather significant experimental effort is presently being mounted in
different laboratories to prove its Majorana character. Here it is the lepton number violating neutrinoless
\dbb\ decay, which is at the center of attention. So far,
the experiments to measure the \dbb\ decay properties have exclusively been focussing on the \bbm\
variant. In these cases one attempts to measure the spectrum of the two emitted electrons, where a discrete line
at the endpoint energy carries the signature of a \znbbm\ decay, and hence, of a Majorana neutrino involved.
Although favored by phase space, the \znbbm\ decay remains elusive because
the decay rate is proportional to the square of the effective neutrino mass, which is the unknown. The decay rate further
contains an unknown and rather complicated nuclear matrix element, which makes precise theoretical predictions of
the decay rate rather involved. Experiments are therefore driven by what one considers the most promising
cases and, at the same time, by what one considers technically most feasible. Experiments that are presently at
a rather advanced stage are those concerned with $^{76}$Ge, $^{100}$Mo and $^{130}$Te ~\bbm\
decays~\cite{Moon,Nemo,Gerda,Klapdor,Cuore}.

On the other hand, it is interesting to note that little attention has been given to alternative decay modes in
\dbb\ decay. I will shortly review, why this is the case rather than go into any of the details of the existing or
proposed experiments. Indeed, it turns out that there may be at least one viable and technically feasible
alternative for measuring the \dbb\ decay in a way that one could prove unambiguously the Majorana
character of the neutrino.

\section{Summary of various \dbb\ decay modes}
\subsection{The \nnbbm\ and \znbbm decay}
It may be instructive to recall some of the leading factors that characterize the decay rate of the \dbb\
decay. In the case of the \bbm\ type, the evaluation  of the \nnbbm\ decay rate is rather
straightforward (assuming a $0^+\rightarrow 0^+$ transition) leading to the simplified
expression~\cite{Prim-Rosen}:

\begin{eqnarray}
\Gamma^{2\nu}_{(\beta^-\beta^-)} &\!\!\!\!=&\!\!\!\! C \left[\frac{G_f}{\sqrt{2}} \cos(\Theta)\right] ^4
\mathcal{F_{\!(-)}}\!\!\!\!\!^2  \;\;|M_{DGT}^{(2\nu)}|^2 \;\; f(Q)\nonumber\\
\mathcal{F_{\!(-)}}&\!\!\!\!=&\!\!\!\!\frac{2\pi\alpha Z}{1-\exp(-2\pi\alpha Z)}
 \label{nnbmbm}
\end{eqnarray}

Here, $G_f$ is the Fermi constant, $\Theta$ is the Cabibbo angle, $\mathcal{F_{\!(-)}}$ is the Coulomb factor
for $\beta^-$ decay, $\alpha$ the fine structure constant and $Z$ the final Z-value of the nucleus. The factor
$C$ is a relativistic correction term for \dbb\ decay, which enhances the decay for high-Z nuclei (C is of
order unity for $Z=20$ and $\sim5$ for $Z=50$). The factor $f(Q)$ can be expressed in terms of a polynomial of
order $Q^{11}$, where $Q$ is the Q-value of the reaction. This high Q-value dependence is essentially a result
of the phase space. The nuclear structure dependence is given by the \dbb\ decay matrix element
$M_{DGT}^{(2\nu)}$. For the case of \nnbbm\ decay this matrix element is the product of two Gamow-Teller
transition operators, which one could readily measure through charge-exchange reactions in the $\beta^+$ and
$\beta^-$ direction at intermediate energies (100-200 MeV/nucleon). There are many recent examples where these
matrix elements have been determined through $(p,n)$ and $(n,p)$ reactions or more recently through $(^3$He$,t)$
and $(d,^2$He) reactions~\cite{frekersca,frekerssn,Grewe-Ca-PRC76,Grewe-Zn64-PRC77,Grewe-Se76-PRC78,Dohmann-Mo96-PRC78,Ejiri}.

The high Q-value dependence makes certain nuclei particularly promising, like for instance $^{48}$Ca with the
largest Q-value in the periodic table of 4.271~MeV or $^{150}$Nd with the second largest one of
3.368~MeV, which comes with  an additional advantageously large Z-value of Z=60. Both have, however, a low isotopic abundance (0.18\% and 5.6\%), which, apart form other technical difficulties, makes them less attractive (though there are new initiatives like CANDLES or SNO$^+$ being put forward). One may also note that the nuclear matrix element $M_{DGT}^{(2\nu)}$
can alter the situation significantly. In the case of $^{48}$Ca it has recently been shown~\cite{frekersca} that
the Gamow-Teller transitions, which excite the intermediate $1^+$ states in $^{48}$Sc  for the $\beta^+$
direction rather strongly, do this only weakly for the $\beta^-$ direction and vice versa. This "state-mismatch"
makes $^{48}$Ca particularly stable. Further, for high-Z nuclei the $\beta^+$ direction of the Gamow-Teller (GT)
operator, which connects the ground state of the daughter with the $1^+$ states of the intermediate nucleus may
be severely Pauli blocked as a consequence of the large (N-Z) value. In how far nuclear deformation at high Z
enters into the matrix elements (like e.g. for $^{150}$Nd) is also still a matter of debate.

The structure of the formula for the \znbbm\ decay is not very much different from Eq.~\ref{nnbmbm}, except that
the square of the effective Majorana neutrino mass $m_\nu^2$ enters into the formula.  The effective neutrino mass is given as:

\begin{equation}
\left\langle m_{\nu _{\rm e}} \right\rangle  = \left|\sum\nolimits_{i} {U_{{\rm e}i}}^{\!\!2} \;m_i\right|
 \end{equation}
 %= \left|\sum\nolimits_{i} {\left| U_{{\rm e}i}  \right|^2} \exp (2i\alpha_{{\rm e}i})m_i \right|

\noindent with $U_{{\rm e}i}$ being the elements of the mixing matrix containing two mixing angles $\theta_{12}$
and $\theta_{13}$ as well as two CP phases $\phi_{12}$ and $\phi_{13}$, and $m_i$ being the three corresponding
mass eigenvalues.

The Q-value dependence is less
pronounced (only $Q^5$), however, the nuclear matrix elements in this case are significantly more complex. Because the Majorana
neutrino appears in the description as a virtual particle with a momentum transfer typically of order $q=0.5
fm^{-1}$, the propagator includes a multitude of virtual transitions in the intermediate nucleus. To get
experimental information about these matrix elements may be an insurmountable task, unless one could show
theoretically, that low lying states of lowest multipolarity (e.g. $1^+$, $2^-$ and $3^+$ states) would be the
strongest contributors to the matrix elements~\cite{SuhonenNPB752}. In these cases, charge-exchange reactions
could be employed to give stronger bounds to the theoretical evaluations of the \znbbm\ decay rates.

Present experimental lower bounds of the half-lives for the \znbbm\ decays are in excess of $10^{24}$y, which would give upper
bounds for the Majorana neutrino mass of significantly less then 1 eV depending on the model for the nuclear matrix elements. There is only one experiment so far, which has claimed evidence for the \znbbm\ decay with a half-life of $1.2\cdot10^{25}\rm y$~\cite{Klapdor}.

\subsection{The \nnbbp\ and \znbbp decays}

\begin{table*}[t]
\caption{Nuclei, which can undergo a \nnbbp\ decay. Note that the energy penalty to be paid equals
$4m_0c^2=2.044$ MeV because of the creation of 2 electron-positron pairs. Of course, also the
$\beta^+\varepsilon$ and the $\varepsilon\varepsilon$ modes are possible, where the energy penalty is either
$2m_0c^2=1.022$ MeV or zero. In these cases an increasing number of final states in the daughter nucleus can be
reached~\cite{iso-Audi}.}
\begin{center}
 \vspace*{1.5mm}
 \tabcolsep 1.8mm
 \renewcommand{\arraystretch}{1.5}
\begin{tabular}{|c|c|c|c|}
  \hline
  transition & Q-value [MeV]& final states in daughter   & isotopic abundance   \\\hline
  $^{78}$Kr $\rightarrow$ $^{78}$Se & 2.864 $\pm$ 0.002 & many   & 0.35\%\\
  $^{96}$Ru $\rightarrow$ $^{96}$Mo & 2.719 $\pm$ 0.008 & many   & 5.52\% \\
  $^{106}$Cd $\rightarrow$ $^{106}$Pd & 2.770 $\pm$ 0.007 & many & 1.25\% \\
  $^{124}$Xe $\rightarrow$ $^{124}$Te & 2.864 $\pm$ 0.002 & many & 0.10\% \\
  $^{130}$Ba $\rightarrow$ $^{130}$Xe & 2.620 $\pm$ 0.003 & many & 0.11\% \\
  $^{136}$Ce $\rightarrow$ $^{136}$Ba & 2.419 $\pm$ 0.013 & many & 0.20\%\\

  \hline
\end{tabular}
\end{center}
\label{2bb-plus} \vspace{-3mm}
\end{table*}

For the \dbb\ decay in the $\beta^+$ direction, there are three alternative processes depending on the
Q-value involved, the $\beta^+\beta^+$, the 1-electron capture version $\beta^+\varepsilon$ and the 2-electron
capture process $\varepsilon\varepsilon$, each either with or without neutrinos. There are only 6 nuclei, which
can undergo $\beta^+\beta^+$ decay. These are listed in Table~\ref{2bb-plus}. All of these have low isotopic
abundance, in most cases even below 1\%. The reason that none of these are attractive candidates for an
experiment is because of a rather severe energy penalty to be paid in $\beta^+\beta^+$ decay, which amounts to
$4m_0c^2 = 2.044$ MeV due to the creation of two electron-positron pairs. This value has to be subtracted from
the Q-values in Table~\ref{2bb-plus} leaving little phase space for the decay. Referring to Eq.~\ref{nnbmbm} the
Q-value dependence has to be replaced by $(Q-4m_0c^2)$, which is rather discouraging given the numbers appearing
in Table~\ref{2bb-plus}. The Coulomb factor $\mathcal{F_{\!(+)}}$ in this case is derived from Eq.~\ref{nnbmbm}
by replacing $Z$ with $-Z$, which makes it even less favorable by another 1 to 2 orders of magnitude. The
effects of the Pauli blocking in the nuclear matrix elements in the 2-neutrino decay are, however, reduced, as
$(N-Z)$ values are smaller and $B(GT)^\pm$ strength values are likely more balanced. Experimentally, one would
be faced with the detection of the kinetic energy of the two emitted positrons together with the detection of
four 511 keV annihilation $\gamma$-rays, which presents a rather serious experimental challenge. However, one
may want to keep in mind that electron capture is an alternative decay mode, which could lead to excited states
in the daughter nucleus, thereby giving a potentially clearer signature of the decay through the accompanied
de-excitation $\gamma$-ray, as will be discussed below.

\subsection{The \nnbec\ and \znbec\ decays}

\begin{table*}[t]
\caption{Nuclei, which can undergo a \nnbec\ decay. Note that the energy penalty to be paid equals
$2m_0c^2=1.022$ MeV because of the creation of one electron-positron pair. Of course, also the
$\varepsilon\varepsilon$ mode is possible, where there is no energy penalty. The interesting case here is the
nucleus \selen, which has a \znecec\ decay Q-value close to the  energy of a $2^{\rm nd}$ excited state in the
daughter nucleus to within 5 keV. Another potentially interesting candidate is $^{112}$Sn, where the $0^+$ level
at 1.871 MeV lies $48.5\pm 4.8$ keV below the Q-value~\cite{iso-Audi}. A $(1S1S)$ atomic hole excitation costs
54.6 keV, which makes the 1.871 MeV transition slightly out of reach given the uncertainties at face value.}

\begin{center}
 \vspace*{1.5mm}
 \tabcolsep 1.2mm
 \renewcommand{\arraystretch}{1.5}
\begin{tabular}{|c|c|c|c|}
  \hline
  transition & Q-value [MeV]& final states in daughter   & isotope abund.  \\\hline
  $^{50}$Cr $\rightarrow$ $^{50}$Ti & 1.167 $\pm$ 0.001 & g.s.    & 4.3\%\\
  $^{58}$Ni $\rightarrow$ $^{58}$Fe & 1.926 $\pm$ 0.001 & g.s., 0.810$(2^+)$, 1.674$(2^+)$  & 68.08\%\\
  $^{64}$Zn $\rightarrow$ $^{64}$Ni & 1.096 $\pm$ 0.001 &  g.s.   & 48.60\% \\
  $^{74}$Se $\rightarrow$ $^{74}$Ge & 1.210 $\pm$ 0.002 & g.s., 0.595$(2^+)$, 1.204$(2^+)$ &  0.89\% \\
  $^{84}$Sr $\rightarrow$ $^{84}$Kr & 1.787 $\pm$ 0.004 & g.s., 0.881$(2^+)$ & 0.56\% \\
  $^{92}$Mo $\rightarrow$ $^{92}$Zr & 1.649 $\pm$ 0.005 & g.s., 0.934$(2^+)$, 1.382$(0^+)$, 1.495$(4^+)$& 14.84\% \\
  $^{102}$Pd $\rightarrow$ $^{102}$Ru & 1.173 $\pm$ 0.004 & g.s., 0.475$(2^+)$, 0.943$(0^+)$, 1.103$(2^+)$, 1.107$(4^+)$& 1.02\% \\
  $^{112}$Sn $\rightarrow$ $^{112}$Cd & 1.9195 $\pm$ 0.005 &\renewcommand{\arraystretch}{0.9}\begin{tabular}{r}

    g.s., 0.617$(2^+)$, 1.224$(0^+)$, 1.312$(2^+)$, 1.416$(4^+)$\\
    1.433$(0^+)$, 1.469$(2^+)$, 1.871$(0^+)$\\
  \end{tabular}
                        & 0.97\%\\
  $^{120}$Te $\rightarrow$ $^{120}$Sn & 1.700 $\pm$ 0.004 & g.s., 1.171$(2^+)$& 0.09\% \\
  $^{144}$Sm $\rightarrow$ $^{144}$Nd & 1.781 $\pm$ 0.004 & g.s., 0.696$(3^-)$, 1.314$(4^+)$, 1.510$(3^-)$, 1.561$(2^+)$& 3.10\% \\
  \hline
  $^{156}$Dy $\rightarrow$ $^{156}$Gd $(*)$ & 2.012 $\pm$ 0.007 & many & 0.06\% \\
  $^{162}$Er $\rightarrow$ $^{162}$Dy $(*)$ & 1.844 $\pm$ 0.004 & many & 0.14\% \\
  $^{168}$Yb $\rightarrow$ $^{168}$Er $(*)$ & 1.422 $\pm$ 0.005 & many & 0.13\%\\
  $^{174}$Hf $\rightarrow$ $^{174}$Yb $(*)$ & 1.103 $\pm$ 0.004 & many & 0.16\%\\
  $^{184}$Os $\rightarrow$ $^{184}$W $(*)$ & 1.451 $\pm$ 0.002 & many & 0.02\%\\
  $^{190}$Pt $\rightarrow$ $^{190}$Os $(*)$  & 1.383 $\pm$ 0.006 &  many  & 0.01\% \\
\hline
\end{tabular}
\end{center}
\label{bec}
\vspace{-5mm}
\begin{flushleft}
\small{$(*)$ At the level of $>10^{20}$ years, $\alpha$-decay will be the dominant decay mode in these cases.}
\end{flushleft}
\end{table*}

Table~\ref{bec} shows all nuclei, which can undergo a $\beta^+\varepsilon$ decay. Here the energy penalty to be
paid is $2m_0c^2=1.022$ MeV, unless the decay proceeds through the 2-electron capture process
($\varepsilon\varepsilon$), which does not require this extra energy. Again, most of the isotopes have low
natural abundances with the exception of $^{58}$Ni, $^{64}$Zn and $^{92}$Mo. An experiment trying to measure the
\nnbec\ and to identify the \znbec\ mode have to cope with the difficulty of how to contain the total visible energy release coming from rather different sources. The electron capture leaves an excited (but neutral) atom behind, which de-excites via X-ray or
Auger-electron emission. In the $0\nu$ case the excess decay energy would have to be carried away as
kinetic energy by the positron, whereas in the $2\nu$ case this energy would be shared by all three light
particles giving in both cases two 511 keV annihilation photons. Further, there usually is rather strong component from the internal Bremsstrahlung electron-capture process (IBEC) producing another extra low-energy photon, which adds to the
complication.

The situation is slightly different for a double-electron capture process in a case, where the
daughter nucleus is left in an excited state. Here the excited state de-excitation provides a unique tag. In the
two-neutrino \nnecec\ mode, the excess energy above the nuclear excitation will be carried away by the
neutrinos, thereby leaving no further signature for the decay (except for the atomic X-rays). In the \znecec\ mode, however, the excess energy
will have to be emitted by an extra radiative process, which, apart from the atomic shell excitations, would produce a
nearly mono-energetic photon. Therefore Table~\ref{bec} also shows, which excited states can be reached in the daughter
nucleus, if the decay were to proceed through the $\varepsilon\varepsilon$ mode only. However, the extra radiative coupling reduces the rate by at least 4 - 5 orders of magnitude.

\begin{figure}[ttt]
\begin{center}
   \includegraphics[scale=0.99]{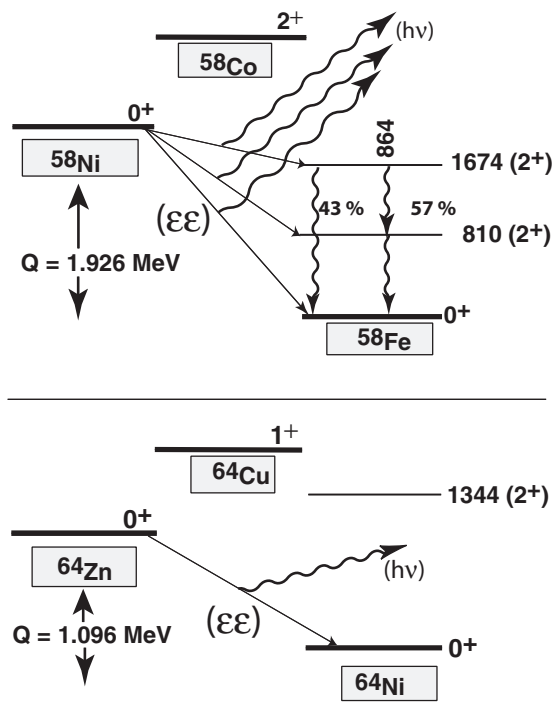}
\vspace*{-0.7cm} \caption{\small Sketch of the various radiative \znecec\ decay modes in $^{58}$Ni and
$^{64}$Zn.} \label{examples}
\end{center}\vspace{-1cm}
\end{figure}

Before turning to the interesting case, \selen, it may be instructive to understand the signatures for two of
the three most abundant isotopes, $^{58}$Ni and $^{64}$Zn in the case of a \znecec\ decay. The situation is
depicted in Fig.~\ref{examples}.

In the \znecec\ decay of $^{58}$Ni, two  excited final states at 1.674 MeV and at 0.810 MeV can be reached,
which in the most fortuitous situation leaves as a signature the 864 keV (branching: 57\%) and 810 keV
$\gamma$-ray cascade as the tag for $^{58}$Fe together with a quasi-monoenergetic radiation of a 252
keV photon. In the less fortuitous case one would still have the 810 keV $\gamma$-ray de-excitation identifying
$^{58}$Fe together with the 1116 keV radiative photon. A similar situation exists for $^{92}$Mo. Less
favorable is $^{64}$Zn, where a \znecec\ process  leaves a single quasi-monoenergetic photon of
1.097 MeV. In Ref.~\cite{Grewe-Zn64-PRC77} the half-life for the \nnecec\ decay with no extra photon was estimated on the basis of experimentally determined nuclear matrix elements, which gave a value of order $1.2\cdot 10^{25}\rm y$. The \znecec\ variant could still be several orders of magnitude longer as a result of the extra radiative coupling bringing it into the range of $10^{30}\rm y$, which is also a value estimated in Ref.~\cite{wycech1}.
Since the double-electron capture process has an $(\alpha Z)^6$ dependence as a result of the atomic
$(1S1S)$-electron density at the nucleus, $^{92}$Mo is favored by an order of magnitude compared to the other
two cases, which could offset its lower abundance. Of course, in all cases the radiative propagator has to be
taken into account, which scales in first order as $(\alpha/Q)^2$, where $Q$ is the energy of the photon.
Further, angular momentum considerations have not been discussed at all, and we note that for a \znecec\ ~$0^+\longrightarrow 0^+$ transition
a radiative $(1S1S)$-capture does not balance angular momentum, unless higher order atomic shell effects are considered as discussed in Ref.~\cite{wycech1}.
%IBEC!!!!!!!!!!!!!!!!!!!!!!!!!!!!!!!

\subsection{The \nnecec\ and \znecec\ decays}

\begin{table*}[t]
\caption{Nuclei, which can only undergo a \nnecec\ or a \znecec\ decay. The decay $^{164}$Er $\rightarrow$
$^{164}$Dy is in itself interesting as the Q-value coincides with atomic levels in the final daughter nucleus,
thereby leaving mostly Auger-electron emission as a de-excitation process in \ecec\ decay~\cite{iso-Audi}). However, $^{164}$Er has an $\alpha$-decay
Q-value of 1.304 MeV, which makes $\alpha$-decay the dominant decay mode and a potential source of background.}
\begin{center}
 \vspace*{1.5mm}
 \tabcolsep 1.8mm
 \renewcommand{\arraystretch}{1.5}
\begin{tabular}{|c|c|c|c|}
  \hline
  transition & Q-value [MeV]& final states in daughter   & isotopic abundance   \\\hline
  $^{36}$Ar $\rightarrow$ $^{36}$S & 0.4325 $\pm$ 0.0002 & g.s.    & 0.34\%\\
  $^{40}$Ca $\rightarrow$ $^{40}$Ar & 0.1936 $\pm$ 0.0002 & g.s.   & 96.94\% \\
  $^{54}$Fe $\rightarrow$ $^{54}$Cr & 0.680 $\pm$ 0.001 & g.s.   & 5.80\% \\
  $^{108}$Cd $\rightarrow$ $^{108}$Pd & 0.272 $\pm$ 0.005 & g.s. & 0.89\% \\
  $^{126}$Xe $\rightarrow$ $^{126}$Te & 0.896 $\pm$ 0.006 & g.s., 0.666$(2^+)$ & 0.09\% \\
  $^{132}$Ba $\rightarrow$ $^{132}$Xe & 0.846 $\pm$ 0.001 & g.s., 0.668$(2^+)$ & 0.10\% \\
  $^{138}$Ce $\rightarrow$ $^{138}$Ba & 0.693 $\pm$ 0.010 & g.s. & 71.70\%\\
  \hline
  $^{152}$Gd $\rightarrow$ $^{152}$Sm $(*)$ & 0.055 $\pm$ 0.004 & g.s.   & 0.20\%\\
  $^{158}$Dy $\rightarrow$ $^{158}$Gd $(*)$ & 0.285 $\pm$ 0.004 & g.s., 0.079$(2^+)$, 0.261$(4^+)$ &  0.10\% \\
  $^{164}$Er $\rightarrow$ $^{164}$Dy $(*)$ & 0.023 $\pm$ 0.004 & g.s. & 1.61\% \\
  $^{180}$W $\rightarrow$ $^{180}$Hf $(*)$  & 0.144 $\pm$ 0.005 &  g.s., 0.093$(2^+)$ & 0.13\% \\
  $^{196}$Hg $\rightarrow$ $^{196}$Pt $(*)$ & 0.821 $\pm$ 0.003  & g.s., 0.355$(2^+)$, 0.689$(2^+)$ & 0.15\%\\

  \hline
\end{tabular}
\end{center}
\label{ecec}
\vspace{-5mm}
\begin{flushleft}
\small{$(*)$ At the level of $>10^{20}$ years, $\alpha$-decay is or will be the dominant decay mode.}
\end{flushleft}
\end{table*}

In Table~\ref{ecec} are listed all nuclei, which for energy reasons can only undergo a double-electron capture (\ecec) process. It
is once again interesting to note that isotopic abundances are low with two marked exceptions, $^{40}$Ca and
$^{138}$Ce. In cases where the Q-values do not allow an excitation of the final nucleus, the \nnecec\ mode
would leave no further trace, except the de-excitation of the atomic shell through  X-ray or  Auger electron
emissions. In the \znecec\ mode the excess energy would have to be radiated away through an extra radiative process as already discussed above.
Since in the $2\nu$ case, radiative processes may compete at the same levels, one would be left with a continuum $\gamma$-ray background below a possible mono-energetic photon line at
the maximum decay energy. The spectrum would be reminiscent of the $\beta^-\beta^-$ decay spectrum, but in a
rather different energy regime. There is a limited number of nuclei ($^{126}$Xe, $^{132}$Ba, $^{158}$Dy,
$^{180}$W and $^{196}$Hg) where excited states can be reached, thereby providing the necessary tag for the
decay; there is only one nucleus ($^{196}$Hg), where a cascade would provide two reasonably high energy
$\gamma$-ray lines for final state identification. However, above mass $A\sim 150$, $\alpha$-decay Q-values are
of order $1-3$ MeV making $\alpha$-decay by far the more dominant decay mode creating an almost irreducible background.

Despite the fact that one can imagine unconquerable experimental difficulties, if one ever wanted to get
involved in such an experimental endeavor, it is nevertheless instructive to enter into a discussion about some
of the properties of the \ecec\ decay.

To estimate the rate of the \znecec\ processes, Eq.~\ref{nnbmbm} has to be modified:

\begin{eqnarray}
\Gamma_{(0\nu\varepsilon\varepsilon)}&\!\!\!\!=&\!\!\!\!\left[\frac{G_f}{\sqrt{2}}
\cos(\Theta)\right] ^4 (M_{DGT} + (\frac{g_V}{g_A})^2 M_{DF})^2\nonumber\\
&&\times|\Psi(1,0)\Psi(2,0)|^2\;\Gamma^\gamma(Q)\;m_\nu^2 .
 \label{znecec}
\end{eqnarray}

Here, $|\Psi(1,0)\Psi(2,0)|^2$ describes the atomic electron density of the first and second electron at the
nucleus. In first order this function scales with $(\alpha Z)^6$, thus, favoring heavy nuclei;
$\Gamma^\gamma(Q)$ describes radiative photon production, and $m_\nu^2$ is the square of the Majorana neutrino
mass. The nuclear physics enters into $M_{DGT}$ and $M_{DF}$ being generalized double Gamow-Teller and
Fermi-operators, which already include the higher order multipoles. Note that  the usual phase space factor does not appear anymore, thus, high rates are expected for the low Q-values, for which the photon propagator
diverges as $(\alpha/Q)^2$. This is in  contrast to the other non-radiative decay modes discussed above. Wycech
and Sujkowski~\cite{wycech1,wycech2} estimated rates for the \znecec\ decays for several of the nuclei in
Table~\ref{ecec} and find half-lives, which are of order $10^{30}$ years. Their calculations do, however,
not contain all effects of the atomic shell. Nevertheless, there is an
interesting effect already noted by these authors. If the Q-value of the \znecec\ decay matches the binding
energy of the atomic shell, a resonant enhancement of the decay rate by several orders of magnitude could be
expected, bringing the half-life down to $10^{25}$y or even below (for a 1 eV mass of the neutrino). The case,
which was discussed, is the $0^+\longrightarrow 0^+$ transition $^{152}$Gd(\znecec)$\longrightarrow ^{152}$Sm,
which features the second lowest Q-value of only 56 keV. This value is close to a (1S2S) atomic two-hole excitation. It was, however, also argued that, if the degeneracy is not complete, then the release of the excess energy requires a magnetic radiative process  to balance the angular momentum~\cite{wycech1}. Atomic de-excitation
would then proceed through X-ray or Auger-electron emission  in the usual way. However, apart from the difficulty to capture any of the
radiation, one would in this case also have to cope with the presence of an $\alpha$-decay, which would be  by many orders of magnitude more intense ($Q_\alpha=2.22$ MeV, $t_{1/2}=1.08\cdot 10^{14}\rm y$). However, the authors also briefly point to a possibility of a transition to an
excited state in the final daughter nucleus. Certainly, one would assume that a Q-value in a \ecec\ process, which  exactly
matches an excited state in the final nucleus added to the atomic shell excitation  would be a coincidence too much
to ask for.

\begin{figure}[ttt]
 \epsfxsize 5.5cm
 \centerline{\epsfbox{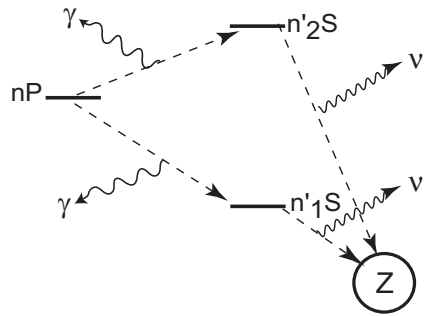}}
    \caption{\small Schematic view of the dominant process in IBEC. A  $nP$-electron emits an E1-photon and populates an intermediate $n'S$-state, from which it is subsequently captured by the nucleus. Figure adapted from Ref.~\cite{deRujula}.}
	\label{fig:feynman}
 \end{figure}

Some remarks on the angular momentum balancing may be in order at this stage. The term \znecec\ hides the fact that in all relevant cases the excess decay energy must be radiated away by an extra photon (i. e. through a $0\nu\gamma\varepsilon\varepsilon$ process  as indicated in Fig.~\ref{examples}). Since an electron-capture process  predominantly involves an $S$ electron with only a few percent contribution from $P_{1/2}$ electrons due to relativistic effects, the  $0\nu\gamma\varepsilon\varepsilon$ process, which creates a double K-shell vacancy,  cannot balance angular momentum in a $0^+\longrightarrow 0^+$ transition, unlike its \nnecec\ counterpart. However, from  the single electron-capture process it is known that it possesses a dominant inner Bremsstrahlung (IBEC) component as a result of the sudden change of charge in the atom. This process renders it a three-body decay with a continuous photon spectrum theoretically extending to an  end-point defined by the Q-value~\cite{deRujula}. The radiative capture process proceeds with a probability of about $10^{-2} - 10^{-4}$ per captured event depending on the total available energy~\cite{Bambynek}. From the view of perturbation theory, IBEC is a second order process, in which both electromagnetic and weak transitions contribute~\cite{Bambynek}. Initial and final states are connected by a set of intermediate (bound) states reminiscent of the two-neutrino nuclear double-beta decay case. The IBEC process is particularly interesting for Q-values, which are close to the typical X-ray energies involved, like in the cases of near degeneracy discussed here. In these cases transitions get greatly enhanced over those that require significantly higher photon energies.  The situation is depicted in Fig.~\ref{fig:feynman}. An $nP$ electron emits E1 radiation and transforms into an intermediate $n'S$ electron. From this state, which has a non-vanishing  electron density  at the origin, the electron weakly interacts with the nucleus and gets captured, thereby leaving a vacancy in the $nP$ state and also making a $0\nu\gamma\varepsilon\varepsilon$  $0^+\longrightarrow 0^+$ an allowed transition. Pronounced resonances occur in the X-ray spectrum at various positions reflecting the difference of the binding energy of the two states involved. This process has been verified in a rather impressive experiment performed by Springer {\it et al.}~\cite{Springer}  on $^{163}$Ho, which features the lowest capture Q-value of $Q\simeq 2.5\rm{keV}$. The IBEC process can, of course, directly be adapted to the case of double-electron capture. A rich number of possible resonant states can be reached in the regime of lowest Q-value, thereby also allowing different angular momentum couplings.

\section{The $^{74}$Se(\znecec)$^{74}$Ge case}

\begin{figure}[tttt]
\begin{center}
   \includegraphics[scale=0.42]{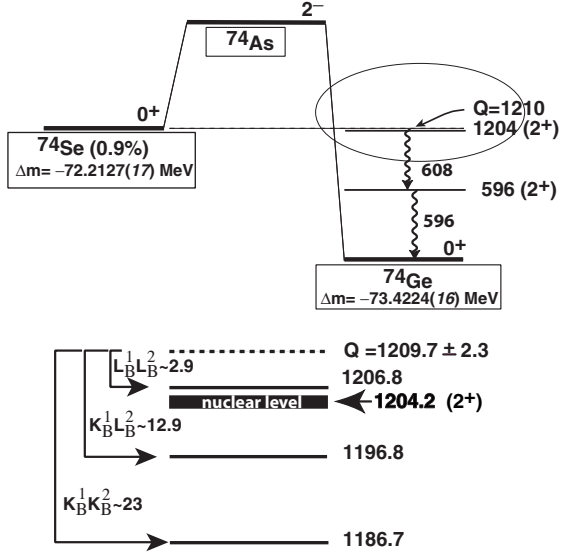}
\vspace*{-0.7cm} \caption{\small Energetics of the \selen\
\znecec\ decay indicating the near degeneracy of the \selen\
ground state and the second excited state in \german. The circle
marks the part, which is magnified in the lower part of the
figure.} \label{selenium}
\end{center}
\end{figure}

As already indicated, the \znecec\ process of \selen\ exhibits  a situation of near degeneracy between the parent and an excited state of the daughter. The Q-value of $1.2097\pm
0.0023$ MeV for the \znecec\ process matches to within 5 keV the second excited state of the daughter at 1.204
MeV, which subsequently de-excites through a 609 keV and 595 keV cascade. Therefore, the mere feeding of this
state would already indicate  the Majorana nature of the neutrino, as there is no phase space left for a
two-neutrino decay mode. The situation gets even more fortuitous as the \nnbec\ variant is also lacking phase
space, because of the energy penalty of $2m_0c^2=1.022$ MeV, which is to be subtracted from the above Q-value.
Thus, there will be no 511 keV annihilation line at any significant level for this decay.

The decay deserves a closer look. In Fig.~\ref{selenium} the energetics of the decay is sketched in detail.
First, it is noted that the Q-value is only known to within an error of 2.3 keV (here: 1.6/1.7 keV error on each
mass), a fact, which is to be kept in mind in the following discussion.

The double-electron capture process leaves the final daughter atom in an excited state with two inner-shell
holes, but otherwise neutral. This inner-shell excitation has to be subtracted from the Q-value. Already for energy reasons, a capture of two
electrons from the K-shell is not allowed. A double K-shell binding energy of $K_B^1K_B^2 \sim 23$ keV would have to be paid,
which places the effective Q-value of the reaction below the $1204$ keV excitation in \german\ near 1186 keV. To
estimate the binding energy one can assume an intermediate charge of $Z=33$ ($^{74}$As) for the first capture
process and $Z=32$ (\german) for the second. Similarly, a K- and an L-shell hole places the effective Q-value at
1196.8 keV, which is  about 7 keV below the nuclear level. Creating a double L-shell vacancy, however,  will cost an
energy of about 2.9 keV, and the effective Q-value now lies 2 keV above the 1204 keV level. This 2 keV value
further coincides with the present precision, with which the mass difference is known, and one may already
speculate about a possible complete energy degeneracy with an atomic final state (including one created through an IBEC process). Such a degeneracy would present a singular opportunity,
as the \znecec\ rate could now be enormously enhanced. Clearly, an effort should be undertaken to re-measure the
masses of \selen\ and \german\ to within a precision of  $\sim 10 \rm {eV}$. This ought to be an easy task using
today's ion trap facilities, like e.g. the one described in Ref.~\cite{Blaum}. Given the estimates of Wycech et
al.~\cite{wycech1,wycech2}, the half-life could well get into a measurable regime, even in a case of a
non-complete degeneracy. A thorough and detailed theoretical treatment of the atomic shell is therefore urgently
warranted. On the experimental side, the detection would require the measurement of the $\gamma$-ray
de-excitation, preferentially in coincidence mode to discriminate against background. The rather low \selen\
isotopic abundance of $0.9\% $ requires an enrichment of rather large quantities. However, the techniques of
measuring extremely low radioactivity levels is well advanced and manageable~\cite{Gerda}.

\section{Conclusion}
The physics of \dbb\ decay has been reviewed, however, focussing on the alternative decays
$\beta^+\beta^+$, $\beta^+\varepsilon$, and the $\varepsilon\varepsilon$. Although the $\varepsilon\varepsilon$
process is generally being viewed as unattractive for an experiment, the \znecec\ decay of \selen\ offers a
remarkable experimental chance to identify the Majorana nature of the neutrino. This is due to an accidental,
almost complete degeneracy of the parent ground state with the second excited state in the daughter nucleus
\german. This degeneracy can enhance the \znecec\ process by
many orders of magnitude. Further, the de-excitation of the \german\ daughter proceeds through an easy to detect
$2\gamma$-ray cascade, which by its mere detection identifies the neutrino as a Majorana particle. A key
ingredient into the rate determination is the exact knowledge of the masses of the two nuclei, \selen\ and
\german. A precision of order 10 eV appears possible using today's ion trap facilities. Such an experiment
should be attempted. The exact rate estimate relies on a rather involved and detailed theoretical treatment of
the decay process, where most of the work appears in the treatment of the atomic shell effects. A theoretical
understanding of these processes may be the first important task, if one ever wanted to consider a large scale
experiment.

\end{document}